\documentclass[twocolumn,showpacs,amsmath,amssymb]{revtex4}
\usepackage{amssymb}
\usepackage{txfonts}
\usepackage{bbm}
\usepackage{graphicx}
\usepackage{appendix}
\usepackage{epsf}
\usepackage{epstopdf}

\begin{document}

\title{Theoretical investigation of the four-layered self-doped high-T$_c$ superconductors: evidence of pair tunneling effect}

\author
{Tao Zhou}

\affiliation{College of Science, Nanjing University of Aeronautics and Astronautics, Nanjing 210016, China}

\date{\today}
\begin{abstract}
Based on a four-layered self-doped $t-J$ type model and the slave-boson mean-field approach, we study theoretically the superconductivity in the electron-doped and
hole-doped layers. The neighbor layers are coupled through both the single electron interlayer hopping and pair tunneling effect. The superconducting gap magnitude for the electron-doped band is nearly twice of that of the hole-doped one, which contrasts to our previous understanding of the electron-hole
asymmetry in high-T$_c$ superconductors but consistent with recent angle-resolved-photoemission-spectroscopy experiments in four-layered materials Ba$_2$Ca$_3$Cu$_4$O$_8$F$_2$. Our results propose that the pair tunneling effect is important to examine the multi-layered superconducting materials.

\end{abstract}
 \maketitle

 The high-T$_c$ superconductivity has not been understood in spite that it has been studied intensively for more than twenty years~\cite{lee}. Generally speaking the materials include the CuO$_2$ layered compound and the CuO$_2$ planes act as the conducting planes. The parent compound is Mott insulator and
 the superconductivity is realized by doping either holes or electrons into the CuO$_2$ planes.

Recently the discovery of the multilayered material Ba$_2$Ca$_3$Cu$_4$O$_8$F$_2$ (F0234) has attracted much attention~\cite{iyo,iyoa}. This compound has four CuO$_2$ planes per unit cell, which can be divided into two
inequivalent groups: the outer layers which have apical F ions and the inner ones. It is expected that this material is a Mott insulator from the valence of the Cu ion obtained by
the canonical chemical formula, while it has been verified by experiments that the compound
is indeed superconducting with the critical temperature 60 K. The angle-resolved-photoemission-spectroscopy (ARPES) experiments
observed  at least two Fermi surfaces~\cite{che,xie,chen}. These two Fermi surface sheets are split along the nodal direction, which is different from the bilayer splitting effect observed in Bi-2212 material, where the two Fermi surfaces degenerate along the nodal direction~\cite{feng}.
In fact, one can conclude from the areas of the Fermi surfaces that the two Fermi surfaces are corresponding to electron-doped and hole-doped ones with the doping around $0.2$~\cite{che}.
This result is consistent with previous study on multilayered systems
that the outer layers have more holes than inner layers~\cite{tok,kot}.
All of the experiments indicate that, due to the presence of the apical F ions, this compound is self-doped with the outer layers and inner layers are hole-doped and electron-doped, respectively.

One of the remarkable features of high-T$_c$ superconducting materials is the electron-hole
asymmetry in the phase diagram, i.e., the superconducting region
is much narrower and the transition temperature is much lower in the electron-doped materials~\cite{taka}. It can be explained based on the Fermi surface topology structure, namely, the hot spots (the crossing points of the Fermi
surface with the magnetic Brillouin zone boundary) for the electron-doped band are close to the nodal point while those of the hole-doped one close to the antinodal point. Thus when we consider the $d$-wave superconductivity meditated by the antiferromagnetic spin fluctuation, the pairing strength for the hole-doped band is stronger~\cite{aba}. On the other hand, the Fermi surface of the hole-doped band is close to the antinodal point, which gives rise to the Van Hove singularities near the Fermi energy. This may be another important ingredient for the electron-hole asymmetry~\cite{mar}.
Since the electron-doped and hole-doped superconducting planes coexist in the F0234 compound,
which may provide an effective way to study in detail the electron-hole asymmetry.
The Fermi surfaces and the corresponding superconducting gaps for the electron-doped and hole-doped bands were observed recently by the ARPES experiments~\cite{che,xie}. The hot spots for the electron-doped Fermi surface are close to the nodal point. And those of hole-doped one is close to antinodal point. These features are similar to those of single-layered materials. Thus one may expect
that the electron-doped band and hole-doped band in F0234 should have similar asymmetric behaviors as those observed previously in single-layered compounds.
However, quite surprisedly, the gap magnitude along the electron-doped Fermi surface is about twice of that along the hole-doped one~\cite{che}. This result contrasts to the asymmetric behavior in single-layer compounds, which suggest that there exists other mechanism in F0234 to be responsible for the electron-hole asymmetry.

Theoretically, the $t-J$ model, with no double occupancy on each site as an additional constraint, is one of the most popular models to study the hole-doped high-T$_c$ superconducting system. The electron-doped system can be mapped to the hole-doped one by taking the electron-hole transformation to satisfy no double occupancy constraint~\cite{toh}. Then the electron-doped and hole-doped systems can be treated in the same manner. For multilayered systems, the inter-layered coupling should be taken into account. The Fermi surface and other physical properties for the bilayer compound were reproduced qualitatively by taking into account a momentum dependent single particle hopping term $t_{\perp {\bf k}}\propto (\cos k_x-\cos k_y)^2$~\cite{nor,jxli}. Similar bilayer model considering one hole-doped layer and one electron-doped layer was also proposed to describe the four-layered compound with the different layers coupled through single electron hopping~\cite{gan} or interlayer Coulomb repulsion~\cite{rib,hoon}. In this way the electron-doped band and hole-doped band were obtained successfully while the unusual electron-hole asymmetry cannot be explained.
The Fermi surface splitting along the nodal direction observed in the experiments cannot be reproduced either~\cite{gan}.
 Thus so far a reasonable explanation for the electron-hole asymmetry and a systematic investigation for the multilayered superconducting systems are still lack.

The motivation of the present work is to fill this void and examine this issue theoretically.
We note that besides the single electron hopping, another important interlayer coupling is the interlayer interaction in the particle-particle channel~\cite{tes} or pair tunneling effect~\cite{whe}. It was also shown that the pair tunneling term is necessary for a reasonable explanation
of the penetration depth in YBCO compound~\cite{xia}. In this paper, the interlayer coupling is considered through both single electron hopping and Josephson-like pair tunneling process. Based on the slave-boson mean-field approach, we show that the experimental results are explained naturally and the pair tunneling term is essential for enhancing the superconductivity.

We start with a Hamiltonian describing a system with four layers per unit cell,

\begin{equation}
H=\sum_l H^{(l)}_p+H_I,
\end{equation}
where $l$ is the layer index with $l=1,4$ representing the hole-doped layers and $l=2,3$ the electron-doped layers, respectively. $H^{(l)}_p$ is expressed by the $t-t^{\prime}-J$ model,
\begin{equation}
H^{(l)}_p=-\sum_{\langle ij \rangle\sigma}t^{(l)} c^{(l)\dagger}_{i\sigma} c^{(l)}_{j\sigma}-\sum_{\langle ij \rangle^{\prime} \sigma}t^{(l)\prime} c^{(l)\dagger}_{i\sigma} c^{(l)}_{j\sigma}-h.c.
+J^{(l)}\sum_{\langle ij \rangle} S^{(l)}_i \cdot S^{(l)}_{j}.
\end{equation}
For the hole-doped layers, $c^{(l)\dagger}_{i\sigma}$ are the creation operators of electrons.
For the electron-doped ones, $c^{(l)\dagger}_{i\sigma}$ are considered as the creation operators of holes to satisfy no double occupancy condition. Then we have $t^{(1,4)}=-t^{(2,3)}=t$, and  $t^{(1,4)\prime}=-t^{(2,3)\prime}=t^{\prime}$.

We use the slave-boson approach to the above in-plane Hamiltonian, namely, the physical particle operators $c^{(l)}_{i\sigma}$
are expressed by slave bosons $b^{(l)}_i$ carrying the charge
and fermions $f^{(l)}_{i\sigma}$ representing the spin, $c^{(l)}_{i\sigma}=b^{(l)\dagger}_i f^{(l)}_{i\sigma}$. At the
mean-field level, the constraint can be written as $b^{(l)\dagger}_i b^{(l)}_i+\sum_{\sigma}f^{(l)\dagger}_{i\sigma}f^{(l)}_{i\sigma}=1$.
In the superconducting state, the pairing operators are defined as $\hat{\Delta}^{(l)}_{ij}= f^{(l)}_{i\uparrow}f^{(l)}_{j\downarrow}-f^{(l)}_{i\downarrow}f^{(l)}_{j\uparrow}$.
We assume the mean-field order parameters, $\Delta^{(l)}_{ ij}=\langle f^{(l)}_{i\uparrow}f^{(l)}_{j\downarrow}-f^{(l)}_{i\downarrow}f^{(l)}_{j\uparrow}\rangle=\pm \Delta^{(l)}$ ($\pm$ are for the bond $\langle ij \rangle$ along $x$ and $y$ direction, respectively) and
$\chi^{(l)}_{ij}=\sum_{\sigma}\langle f^{(l)\dagger}_{i\sigma}f^{(l)}_{j\sigma}\rangle=\chi^{(l)}_0$.  The boson operators $b^{(l)}_i$ condense in the superconducting state, $b^{(l)}_i=b^{(l)\dagger}_i=\sqrt{\delta}$, where $\delta$ is the doping density .

The $H_I$ term is the inter-layer coupling, including the single particle hoping term $H_{IS}$ and pair tunneling term $H_{IP}$,
\begin{eqnarray}
H_I=H_{IS}+H_{IP}
\end{eqnarray}
\begin{eqnarray}
H_{IS}&=&-\delta t_\perp \sum_{i} f^{(1)\dagger}_{i\sigma} f^{(2)\dagger}_{i\sigma}-\delta t_\perp \sum_{i} f^{(2)}_{i\sigma} f^{(3)\dagger}_{i\sigma}
\nonumber\\&&-\delta t_\perp \sum_{i} f^{(4)\dagger}_{i\sigma} f^{(3)\dagger}_{i\sigma}+h.c.
\end{eqnarray}
\begin{eqnarray}
H_{IP}&=&-\lambda \sum_{\langle ij\rangle} \hat{\Delta}^{(1)\dagger}_{ij} \hat{\Delta}^{(2)\dagger}_{ij}-\lambda \sum_{\langle ij\rangle} \hat{\Delta}^{(2)}_{ij} \hat{\Delta}^{(3)\dagger}_{ij}
\nonumber\\&&-\lambda \sum_{\langle ij\rangle} \hat{\Delta}^{(4)\dagger}_{ij} \hat{\Delta}^{(3)\dagger}_{ij}+h.c.
\end{eqnarray}

Then the mean-field Hamiltonian in the momentum space can be written as,
\begin{eqnarray}
H=\sum_{{\bf k}l\sigma}\varepsilon^{(l)}_{\bf k}f^{(l)\dagger}_{{\bf k}\sigma}f^{(l)}_{{\bf k}\sigma}+\sum_{l\bf k}\Delta^{(l)}_{\bf k}
(f^{(l)\dagger}_{{\bf k}\uparrow}f^{(l)\dagger}_{-{\bf k}\downarrow}+h.c.)\nonumber\\- \sum_{k\sigma}[t_{\perp \bf k}(f^{(1)\dagger}_{\bf k\sigma}f^{(2)\dagger}_{\bf k\sigma}+f^{(2)}_{\bf k\sigma}f^{(3)\dagger}_{{\bf k}\sigma}
+f^{(4)\dagger}_{\bf k\sigma}f^{(3)\dagger}_{\bf k\sigma}+h.c.)]+\varepsilon_0,
\end{eqnarray}
where $\varepsilon^{(l)}_{\bf k}=-(2\delta t^{(l)}+J^{(l)}\chi^{(l)}_0)(\cos k_x+\cos k_y)-4\delta t^{(l)\prime}\cos k_x \cos k_y-\mu^{(l)}$, $\Delta^{(1)}_{\bf k}=\Delta^{(4)}_{\bf k}=\widetilde{\Delta}^{(1)}(\cos k_x-\cos k_y)$,
$\Delta^{(2)}_{\bf k}=\Delta^{(3)}_{\bf k}=\widetilde{\Delta}^{(2)}(\cos k_x-\cos k_y)$,
$t_{\perp \bf k}=\delta t_\perp(\cos k_x-\cos k_y)^2/4$, and $\varepsilon_0=2N\sum_l J^{(l)}(\chi^{(l)2}_0+\Delta^{(l)2}/2)+4N\lambda (2\Delta^{(1)}\Delta^{(2)}+\Delta^{(2)2})$.
The renormalized gap magnitudes for the hole- and electron-doped layers are expressed as, $\widetilde{\Delta}^{(1)}=J^{(1)}\Delta^{(1)}+2\lambda \Delta^{(2)}$ and $\widetilde{\Delta}^{(2)}=J^{(2)}\Delta^{(2)}+2\lambda (\Delta^{(1)}+\Delta^{(2)})$, respectively.

Diagonalizing the above Hamiltonian [Eq.(6)], one can obtain four energy bands, expressed by $\zeta_{{\bf k}\psi \nu}=\varepsilon_{-\bf k}+(\psi t_{\perp\bf k}+\nu\sqrt{c_{\bf k}-\psi d_{\bf k}})/2$, where
$\psi,\nu=\pm 1$ and we define $\varepsilon_{\pm\bf k}=(\varepsilon_{\bf k}^{(1)}\pm\varepsilon_{\bf k}^{(2)})/2$, $c_{\bf k}=4 \varepsilon_{+\bf k}^{2}+5t_{\perp\bf k}^{2}$ and $d_{\bf k}=4\varepsilon_{+\bf k}t_{\perp\bf k}$. The energy bands for $\nu=+1 (-1)$ represent the hole-doped (electron-doped) bands, respectively. Then the energy bands in the superconducting state can be obtained, $E_{{\bf k}\tau\psi\nu}=\tau\sqrt{\zeta_{{\bf k}\psi \nu}^{2}+\Delta^{(l)2}_{\bf k}}$, where $\tau=\pm 1$, and the layer $l=1 (2)$ for $\nu=+1 (-1)$, respectively.

The free-energy is written as,
\begin{equation}
F=-2T \sum_{{\bf k}\tau\psi\nu}\ln [2\cosh (\frac{E_{\bf k\tau\psi\nu}}{2T})]-(\mu^{(1)}+\mu^{(2)})N+\varepsilon_0.
\end{equation}
The mean-field parameters $\chi^{(l)}_0$, $\Delta^{(l)}$, and the chemical potentials $\mu^{(l)}$ are calculated from the self-consistent equations obtained by $\partial F/\partial \chi^{(l)}_0=0$, $\partial F/\partial \Delta^{(l)}=0$, and
$\partial F/ \partial \mu^{(l)}=-(1-\delta)N$. We set $t^{(1,4)}=-t^{(2,3)}=1$, $t^{(1,4)\prime}=-t^{(2,3\prime)}=-0.3$, $J^{(l)}=0.3$, and the temperature $T=0.0005$.

The self-doped behavior indicates the charge imbalance between the hole-doped layers and electron-doped ones.
We define the imbalance constant $W=-(\mu^{(1)}+\mu^{(2)})/2$. The system is half-filled when the imbalance constant is zero and self-doped for non-zero imbalance constant. The imbalance constant as a function of the doping $\delta$ is plotted in Fig.1(a). As seen, for all the parameters we considered, $W$ equals to zero for zero doping. And $\mid W\mid$ increases monotonously as the doping increases.

\begin{figure}
\centering
  \includegraphics[width=7cm]{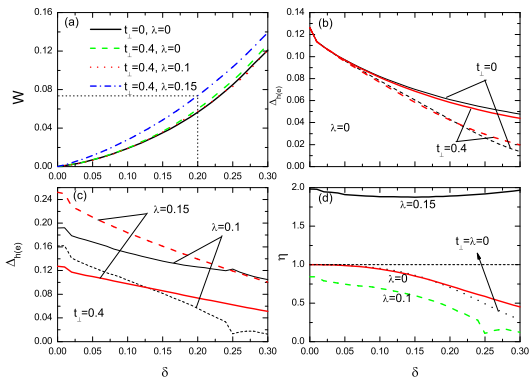}
\caption{(Color online) (a) The charge imbalance versus the doping $\delta$. (b) The gap magnitude as a function of doping when the pair tunneling effect is absent. The solid lines and dashed lines are for the hole-doped and electron-doped cases, respectively. (c) The same as panel (b) but with both single particle hopping and the pair tunneling effect. (d) The gap ratios (see text) with different pair tunneling constants.  }
\end{figure}

The gap magnitudes as a function of the doping $\delta$ without the pair tunneling effect are plotted in Fig.1(b). As $t_\perp=\lambda=0$, corresponding to the single-layer superconductors, the gap magnitude for the hold-doped system is larger than that of electron-doped one. This electron-hole asymmetry is consistent with previous experimental and theoretical results. When the inter-layer single particle hopping term is considered, as shown, the gap magnitudes for the both bands change slightly because the Fermi surface topology changes slightly.
While the gap magnitude for the hole-doped layer is always larger than that of the electron-doped one even for a very large single-particle hopping constant, which indicates that the single particle hopping term cannot explain the abnormal electron-hole symmetry in the F0234 sample.

The gap magnitudes taking into account both the pair tunneling term and the single particle hopping term are plotted in Fig.1(c). As the pair tunneling is weak ($\lambda=0.1$), the gap magnitude of the hole-doped layer is still larger and the asymmetry appears for various doping densities. However, as the pair tunneling constant $\lambda$ increases, the asymmetry is reversed, namely, for the case of $\lambda=0.15$, the gap magnitude for the electron-doped layer is larger than that of the hole-doped one. This is consistent with the experimental observations~\cite{che} and our results propose that the pair tunneling effect may account for the abnormal electron-hole asymmetry in the F0234 samples.

Our results are summarized in Fig.1(d). We define the ratio $\eta=\widetilde{\Delta}^{(2)}/\widetilde{\Delta}^{(1)}$. As seen, for the cases where the pair tunneling constant is absent or quite weak, $\eta$ is less than one. This asymmetry is due to the Fermi surface topology which has been discussed intensively in previous studies. As the pair tunneling constant increases to 0.15, as seen, the gap ratio reaches near $2.0$.  Interestingly, this ratio does not increase as $\lambda$ increases further (not shown here). We have checked numerically that the ratio is around $1.6\sim 2.0$ even for very large $\lambda$.
Our results agree well with the experimental observation, which gives the ratio around $2.0$~\cite{che}.

The origin of the described anomalous asymmetry can be elaborated at the mean-field level, i.e., from the mean-field Hamiltonian [Eq.(6)], the pair tunneling term can be absorbed to the pairing potentials and increases the effective potentials. For the electron-doped layer there are two neighbor layers thus the effective pairing potential is larger than that of the hole-doped one which has only one neighbor layer. As a result, the gap magnitude of the electron-doped layer is larger than that of the hole-doped one as the pair tunneling constant increases.
Our results propose that the pair tunneling effect should play an important role when one investigates the physical properties of the multi-band superconductors.

\begin{figure}
\centering
  \includegraphics[width=6cm]{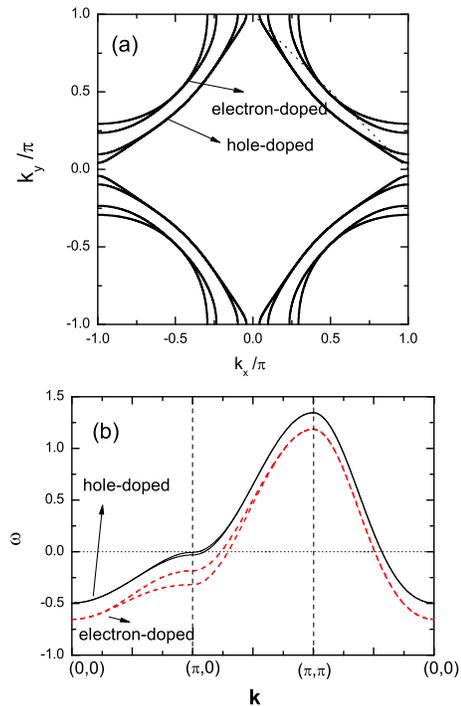}
\caption{(Color online) The normal state Fermi surface (a) and band structure (b) with $t_\perp=0.4$ and $\lambda=0.15$.  }
\end{figure}

We now study the normal state Fermi surface.
The parameters are fixed as $t_\perp=0.4$, $\lambda=0.15$, and the doping $\delta=0.2$ [corresponding to the imbalance constant
$W=0.0735$ displayed in Fig.1(a)] in the following presented results.
 As seen in Fig.2(a), there are four sheets of Fermi surface, among which two being electron-doped type with hot spots close to the nodal point and two hole-doped type with hot spots close to the antinodal point. The two distinct bands near the nodal direction are observed. Each band splits into two when closer to the antinodal direction. Our result of the Fermi surface near the nodal direction is consistent with the ARPES experiments~\cite{che,xie}. The band splitting near the antinodal direction is consistent with LDA prediction~\cite{xie}. While the Fermi surface revealed by the experiments is not distinct enough and large error bars exist~\cite{che,xie}. Thus the band splitting effect near the antinodal direction has not been resolved by experiments yet and still needs further verification.
The band structure along the path $(0,0)\rightarrow (\pi,0) \rightarrow (\pi,\pi) \rightarrow (0,0)$ is plotted in Fig.2(b). As seen, there is one electron-doped band and one hole-doped band near $(0,0)$, respectively. We can also see that both bands spilt and the flat regions appear near the antinodal point $(\pi,0)$ indicating the Van Hove singularities in the density of
states. For the hole-doped bands the flat regions appear near the Fermi energy and the splitting energy is small. For the electron-doped ones the flat regions are at the higher energy with the splitting energy is larger. The spilt bands combine into one as the momentum moves away from $(\pi,0)$ towards the diagonal direction.

\begin{figure}
\centering
  \includegraphics[width=6cm]{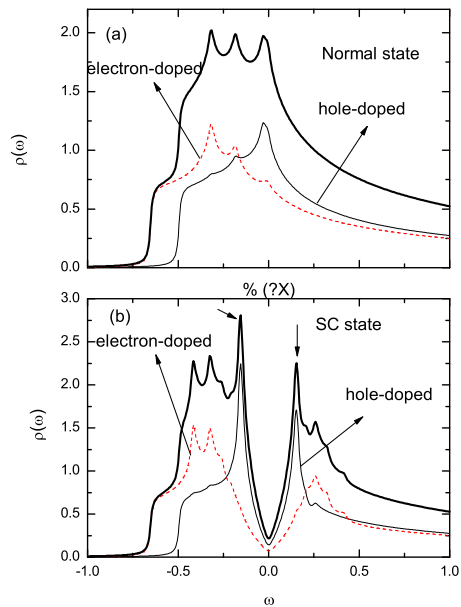}
\caption{(Color online) The density of states in the normal state and superconducting state with $t_\perp=0.4$ and $\lambda=0.15$. The (blue) solid and (red) dashed thin lines are the density of states contributed from hole-doped and electron-doped bands, respectively. }
\end{figure}

At last let us study the density of states spectra in the normal state and superconducting state, respectively. In the normal state, as seen in Fig.3(a), there are several peaks shown up at the negative energy, corresponding to the Van Hove singularities as we have mentioned above. For the hole doped band, because the band splitting is small from the band structure shown in Fig.2(b), thus there is one peak near the Fermi energy.
The two peaks at lower energies are contributed by the electron-doped band. In the superconducting state, as seen in Fig.3(b), we can observe clearly the superconducting coherent peaks of the hole bands at the lower energy and those of the electron bands at higher energy. For the whole density of states, the sharp superconducting coherent peaks contributed by the hole-doped bands are seen clearly (labeled by the
arrows). The low energy Van Hove peak shown in the normal state is covered up by the superconducting gap. While for the electron-doped bands, the superconducting peaks are broad and renormalized by the Van Hove peaks. As a result, the whole density of states includes two superconducting coherent peaks caused by the hole-doped band and several broad peaks contributed by the electron-doped bands, including the superconducting peaks and the negative energy Van Hove peaks, respectively.

In summary, we have examined the superconductivity of a four-layered self-doped system based on the $t-J$ type model and the slave-boson mean-field approach. Both single particle hopping and the pair tunneling effect have been taken into account to describe the inter-layer coupling. The electron-hole asymmetry is obtained for the single-layered case and the result is consistent with previous experimental and theoretical results. For the case of the four-layered model, similar asymmetric behavior is obtained when only the single particle hoping is assumed. While the asymmetry behavior is reversed when the pair tunneling effect is added, namely, the gap magnitude for the electron-doped band is nearly twice of that for hole-doped band. This result is consistent with the experimental observation and our results propose that the pair tunneling effect should be an important issue to study the superconductivity of multi-layered materials.

This work was supported by the NSFC under the Grant No. 11004105 and Scientific Research Foundation from Nanjing University of Aeronautics and Astronautics.

\end{document}